\documentclass[prd,nofootinbib,showpacs,rotate]{revtex4}

\usepackage{graphicx}
\usepackage{dcolumn}
\usepackage{bm}
\usepackage{color}
\input{epsf}
\begin{document}

\title{  Static strings in Randall-Sundrum scenarios and the quark anti-quark potential: Erratum}

\author{Henrique Boschi-Filho,  Nelson R. F. Braga  and Cristine N. Ferreira}


\begin{abstract} 
We correct the energy of the static strings in hep-th/0512295,  for large quark anti-quark separation.
This energy is a smooth function of the quark separation for any position of the infrared brane.
The asymptotic behavior of this energy is that of the Cornell potential as stated in the article.
However, this identification does not fixes the AdS radius.

\end{abstract}

\maketitle
The lines that we considered as the geodesics in the article \cite{Boschi-Filho:2005mw} are not
the correct ones for $L  > L_{crit} $. This is so because  they do not generate the minimum world sheet area. 
The correct geodesics in this case correspond to the two halves of the curve at $ L =  L_{crit} $, split in the middle point and connected by a straight line along the brane at $r=r_2$.
Note that the shape of the curved parts of the geodesics do not vary with $L$ for $L  > L_{crit} $, in contrast to what was considered in the article.
Equation (8) must be replaced by
\begin{eqnarray*}
E_{RS}^{\,(+)} &=& \frac{ r_2 }{\pi \alpha^\prime} 
\Big[ I_2 (r_1/r_2) - 1 \,\Big]
\,+\, \frac{ r_2^2 }{2 \pi \alpha^\prime\,R^2}\,( L - L_{crit})\,  \\
&=& \frac{ r_2 }{\pi \alpha^\prime} 
\Big[ I_2 (r_1/r_2) \,-\, I_1 (r_1/r_2 )    - 1 \,\Big]
\,+\,   \frac{ r_2^2 }{2 \pi \alpha^\prime\,R^2}\,L\,\,,\hskip 3cm (8)
\end{eqnarray*}

\noindent where we used the definition of $L_{crit}$ given by equation (4) with $r_0 = r_2$. 
According to the definition of the Randall Sundrum space the quark brane is located at $r_1 = R$ in equation (8).  This energy is a smooth function of the parameter $L$, in contrast to what is said in the article 
(including the abstract). So, figure 2 is wrong and must be disconsidered. Actually, even the expression (8) in the article does not lead to a discontinuity in the derivative of the energy with respect to $L$. 

Since there is no discontinuity in the energy derivative it is not necessary to fix the infrared (IR) brane position $r_2$. The identification $r_2 = R$ has to be understood just as a particular choice which is consistently made in the rest of the article. 

Equations (12) and (13) in the article are wrong and must be replaced by 
$$
E^{\,(+)}\,=\, \frac{ R }{\pi \alpha^\prime} 
\Big[ I_2 (r_1/R) - I_1 (r_1/R ) - 1 \,\Big]
\,+\,\frac{ 1 }{2 \pi \alpha^\prime\,}\, L \,  \,, 
 \hskip 5cm (12) 
$$
\noindent that corresponds to the new equation (8) with quark position $r_1 > R $ and IR brane position
 $r_2 = R$.
This is the correct static string energy for $ L \ge L_{crit} $. 
This energy has the same asymptotic behavior as the incorrect expression presented in the article.
That means, it behaves asymptotically as the heavy quark anti-quark Cornell potential. 
From equations (11) and  (12), using the Cornell parameters $a$ and $\sigma$ (with the choice $r_2 = R$ ), one finds that equation (15) must be replaced by 
\begin{eqnarray} 
E &=&
\left\{ \matrix { \displaystyle \frac{ 4 a }{ 3 C_1^2} \,\frac{I_1 (r_1/r_0 )}{L}
\Big[ \, I_2 (r_1/r_0)\,-\,1 \Big]\,, \hskip 3cm  L \le L_{crit} \cr \cr 
\displaystyle \sqrt{\frac{4 a \sigma}{3 C_1^2}}\,\,
\Big[ I_2 (r_1/R) - I_1 (r_1/R ) - 1 \,\Big]
\,+\,\sigma\, L \,
 \hskip 3cm   L \ge L_{crit}
} \right. \,,\,\,\,\,\,(15)\nonumber 
\end{eqnarray}
   
\noindent where $ L_{crit}\,=\, 2R I_1 (r_1/R )$.
When we take the quark position ($r_1$) going to infinity, equation (16) must be replaced by 
\begin{eqnarray} 
E &=&
\left\{ \matrix { \displaystyle - \frac{ 4 a }{ 3 L }
, \hskip 5cm  L \le L_{crit}  \cr \cr 
\displaystyle -\, 4\, \sqrt{\frac{ a \sigma}{3 }}
\,+\,\sigma\, L \,
 \hskip 3cm   L \ge L_{crit} 
} \right. \,\,\,\,\,\,(16)\nonumber 
\end{eqnarray}

\noindent where $\,L_{crit}\,= \,2RC_1\,$. 
The identification of the static string energy in our model (without choosing a value for $r_2 $) 
with the Cornell potential leads to the relations 
$$ 2 \pi \alpha^\prime\, R^2 \,\sigma \, =\,  r_2^2  \,\,\,\,\,\,\,,\,\,\,\,\,\,\,\, 
2 \pi \alpha^\prime \, a \,=\,  3 C_1^2 R^2 \,\,.$$

\noindent That means: the identification with the Cornell potential does not imply  
a fixed value for the AdS radius $R$. 
Equation (17) of the article has to be understood just as an 
effective value for $R$  corresponding to the particular choice: $r_2 = R$.


\begin{thebibliography}{99}

% \cite{Boschi-Filho:2005mw}
\bibitem{Boschi-Filho:2005mw}
  H.~Boschi-Filho, N.~R.~F.~Braga and C.~N.~Ferreira,
  %``Static strings in Randall-Sundrum scenarios and the quark anti-quark
  %potential,''
  Phys.\ Rev.\ D {\bf 73} (2006) 106006, 
 [arXiv:hep-th/0512295].
  %%CITATION = HEP-TH 0512295;%%

\end{thebibliography}
\end{document}